\documentclass[a4paper,10pt]{article}
\usepackage{graphicx}
\usepackage[labelfont={it,bf},font=it]{caption}
\usepackage{amsmath}
\usepackage{amsfonts}
\usepackage{pslatex}

\makeatletter
\renewcommand\section{\@startsection {section}{1}{\z@}%
    {-3.5ex \@plus -1ex \@minus -.2ex}%
    {2.3ex \@plus.2ex}%
    {\normalfont\bfseries\MakeUppercase}}
\makeatother

\begin{document}

\pagestyle{empty}

\begin{center}
{\fontsize{16}{16pt} \bf
Adding SALT to Coupled Microcavities \\
the making of active photonic molecule lasers
}
\vspace{12pt}\vspace{12pt}\\
{\bf Denis Gagnon, Joey Dumont, Jean-Luc D\'eziel and Louis J. Dub\'e*}\\
{\it D\'epartement de Physique, de G\'enie Physique, et d'Optique, Facult\'e des Sciences et de G\'enie \\
Universit\'e Laval, Qu\'ebec, QC G1V 0A6, Canada\\
*Corresponding author: ljd@phy.ulaval.ca}
\end{center}
\vspace{-0.6cm} \vspace{6pt}

\section*{Abstract}
A large body of work has accumulated over the years in the study of the optical properties of
single and coupled microcavities for a variety of applications, ranging from filters to sensors and
lasers. The focus has been mostly on the geometry of individual resonators and/or on their
combination in arrangements often referred to as photonic molecules (PMs). 

Our primary concern will be the lasing properties of PMs as ideal candidates for the fabrication
of integrated microlasers, photonic molecule lasers. Whereas most calculations on PM lasers
have been based on cold-cavity (passive) modes, i.e. quasi-bound states, a recently formulated
steady-state ab initio laser theory (SALT) offers the possibility to take into account the
spectral properties of the underlying gain transition, its position and linewidth, as well as
incorporating an arbitrary pump profile. We will combine two theoretical approaches to characterize the lasing properties of PM lasers:
for two-dimensional systems, the generalized Lorenz-Mie theory will obtain the resonant modes
of the coupled molecules in an active medium described by SALT. Not only is then the
theoretical description more complete, the use of an active medium provides additional
parameters to control, engineer and harness the lasing properties of PM lasers for ultra-low
threshold and directional single-mode emission.

\section{Introduction}

Harnessing the optical properties of
photonic atoms and molecules allows for a wide variety of applications including metrology, filters and biosensors \cite{Boriskina2010}. This contribution is mainly concerned with the application of coupled microcavities for laser applications, forming what we refer to as a \emph{photonic molecule laser}. The two most important design objectives of two-dimensional (2D) microcavity lasers are low threshold and highly directional emission \cite{Harayama2011}. The traditional way to numerically predict the lasing thresholds and directionality of a given PM geometry is to compute its quasi-bound (QB) modes, defined as solutions of the 2D Helmholtz equation
\begin{equation}\label{eq:helm}
[\nabla^2 + \epsilon (\mathbf{r}) k^2]\varphi(\mathbf{r}) = 0.
\end{equation}
The QB states are sometimes referred to as \emph{cold-cavity} modes since they are often used to compute the emission characteristics of microcavities in the absence of gain. Although they provide a qualitatively useful picture of laser emission, QB modes cannot accurately describe the steady-state lasing behavior of an array of active photonic atoms, even near threshold \cite{Ge2010}. This is due to the QB eigen-frequencies being complex everywhere outside the atoms, resulting in exponential growth of the electromagnetic energy at infinity \cite{Ge2010}. To improve the description of lasing modes, researchers have developed a \emph{steady-state ab initio laser theory} (SALT) \cite{Ge2010th}. The most important feature of SALT is the introduction of a new kind of eigenstate called a constant-flux (CF) state \cite{Harayama2011}. CF states are parametrized by real wavenumbers outside the so-called \emph{cavity region} (corresponding to the region of space filled with active medium) and are thus physically meaningful \cite{Harayama2011, Ge2010th}. In other terms, SALT is a stationary formulation of the Maxwell-Bloch theory that explicitly takes the gain medium parameters into account.

In a previous contribution, we have used SALT to highlight the fact that the lasing thresholds of PM lasers may be strongly affected by the underlying gain medium parameters, specifically the gain transition frequency and linewidth \cite{Gagnon2014}. In this conference paper, we address a related issue, that is the importance of using SALT to characterize the emission directionality of PMs.

\section{Constant-flux states of photonic molecules}
The basis of CF states satisfy the following modified Helmholtz equation
\begin{subequations}
\begin{align}\label{eq:cf}
[\nabla^2 + \varepsilon(\mathbf{r}) K^2(k)]\varphi &= 0, \qquad \mathbf{r} \in C \\
[\nabla^2 + \varepsilon(\mathbf{r}) k^2]\varphi &= 0, \qquad \mathbf{r} \notin C
\end{align}
\end{subequations}
where $C$ is the cavity region. For the purpose of this contribution, we suppose that every photonic atom is active, in other terms contained in $C$. The eigenvalues $K$ are complex and depend on $k$, the \emph{external frequency} (a real number). This formulation ensures that the total electromagnetic flux outside the cavity is conserved \cite{Ge2010th}.

Photonic molecules composed of coupled cylinders can be modeled in a straightforward way using a 2D Generalized Lorenz-Mie Theory (2D-GLMT), also called multipole method \cite{Andreasen2011}. Recently, we have used the combination of SALT and 2D-GLMT to compute the lasing frequencies, thresholds and field distributions of a PMs composed of two coupled cylinders of slighty different radii \cite{Gagnon2014}. For completeness, we review the main equations of the method. Consider an array of $N$ cylinders of radii $u_n$ and relative permittivity $\epsilon_n$. Let also $\mathbf{r}_n = (\rho_n,\theta_n)$ be the cylindrical coordinate system local to the $n^{th}$ cylinder. Furthermore, we will assume that every cylinder is infinite along the axial $z$ direction. The central hypothesis of 2D-GLMT is that the field outside the cylinders can be expanded in a basis of cylindrical functions centered on each individual cylinder, that is
\begin{equation}\label{eq:scat}
\varphi(\mathbf{r}_n) = \sum_{n=1}^N \sum_{l'=-\infty}^{\infty} b_{nl'} H^{(+)}_{l'} (k_0 \rho_{n}) e^{il'\theta_n}
\end{equation}
where $H^{(+)}_l$ is a Hankel function of the first kind. Using Graf's addition formula for cylindrical functions, the characteristic equation of the eigenstates of the cylinder array can be cast as
\begin{equation}\label{eq:characteristic}
 \det[ \mathbf{T}(k_n,k_0)] = 0
\end{equation}
where $k_n$ is the frequency inside the $n^{th}$ cylinder and $k_0$ is the exterior frequency (both can be complex). The transfer matrix $T$ is composed of blocks containing coupling coefficients between cylindrical harmonics centered on each circular cylinder. Its elements are given by 
\begin{equation}
\mathbf{T}_{nn'}^{ll'}  (k_n,k_0)= \delta_{nn'}\delta_{ll'} -(1 - \delta_{nn'}) e^{i(l'-l) \phi_{n'n}} H^{(+)}_{l-l'}(k_0 R_{nn'})s_{nl} (k_n,k_0)
\end{equation}
where $R_{nn'}$ is the center-to-center distance between cylinders $n$ and $n'$ and $\phi_{n'n}$ is the angular position of cylinder $n'$ in the frame of reference of cylinder $n$. The $s_{nl}$ factor results from the application of electromagnetic boundary conditions and is given by
\begin{equation}\label{eq:snl}
s_{nl}(k_n, k_0) = -\dfrac{J_l'(k_0 u_n) - \Gamma_{nl} J_l (k_0 u_n)}{H^{(+)\prime}_l(k_0 u_n) - \Gamma_{nl} H^{(+)}_l (k_0 u_n)}
\end{equation}
where
\begin{equation}
\Gamma_{nl} = \xi_{n0} \dfrac{k_n J_l'(k_n u_n)}{k_0 J_l(k_n u_n)}
\end{equation}
and $\xi_{ij} = 1 \left( \epsilon_j / \epsilon_i \right)$ for TM (TE) polarization. 
Prime symbols indicate differentiation with respect to the whole argument. In the case of an $N$-atomic PM, the matrix $\textbf{T}$ is composed of $N \times N$ block matrices, where the size of the blocks is chosen sufficiently large to ensure convergence of the cylindrical function expansions. More details on the method can be found in \cite{Gagnon2014, Nojima2005, Andreasen2011, Gagnon2012a}.

With this formulation, it is straightforward to compute and compare the QB and CF states of a PM simply by substituting appropriate values of $k_n,k_0$ in \eqref{eq:characteristic}. In the case of QB states, the appropriate substitution is $k_n \rightarrow  k \sqrt{\epsilon_n}$ and $k_0 \rightarrow  k \sqrt{\epsilon_0}$, with complex $k$. As for the CF states, the appropriate substitution is $k_n \leftarrow K \sqrt{\epsilon_n}$ with complex $K$ and $k_0 \rightarrow  k \sqrt{\epsilon_0}$ with complex $k$. One then looks for solutions in the complex $k$ and $K$ plane, respectively. Moreover, each QB state can be uniquely mapped to a unique CF state exhibiting the same symmetries.

Once the eigenmodes are obtained using a root-finding method, one can compute the far-field emission profile in a straightforward fashion. In the domain outside the circle containing all cylinders, eq. \eqref{eq:scat} can be recast in a global frame of reference as \cite{Nojima2005}
\begin{equation}
\varphi(r,\theta) = \sum_n \sum_{l,l'=-\infty}^{\infty} b_{nl'} H^{(+)}_{l}(k_0 r) J_{l-l'}(k_0 R_n) e^{il\theta} e^{i(l'-l)\phi_n}
\end{equation}
where $R_{n}$ is the distance between the center of the cylinder $n$ and the origin of the global coordinate system, and $\phi_n$ is the angular position of cylinder $n$ in that same frame of reference. Using the Sommerfeld radiation condition, the field at $r \rightarrow \infty$ can be written as
\begin{equation}
\varphi(r,\theta) = g(\theta ) \frac{e^{ik_0r} }{\sqrt{k_0 r}}.
\end{equation}
Using the asymptotic expansion of cylindrical functions, one finally obtains the following far-field distribution
\begin{equation}
g(\theta) = \sqrt{\frac{2}{\pi}} e^{i\pi/4} \sum_n \sum_{l,l'=-\infty}^{\infty}
b_{nl'} J_{l-l'}(k_0 R_n) \exp \left[il \left(\theta- \frac{\pi}{2} \right) + i (l' -l) \phi_n \right].
\end{equation}

\begin{figure}
\centering
\includegraphics[]{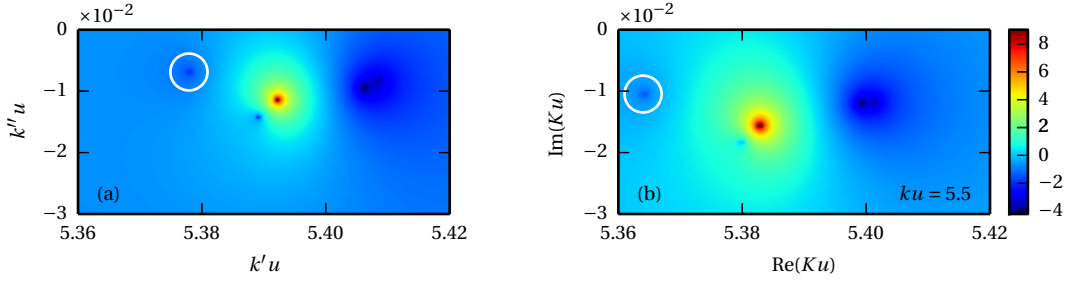}
\caption{Evolution of $\log|\det[\mathbf{T}]|$ in the complex $k$ plane (QB states, left panel) and the complex $K(k)$ plane (CF states, right panel). The signature of four eigenmodes of the triatomic PM can be seen as zeros of the determinant of the transfer matrix.}\label{fig:detmap}
\end{figure}

\section{Case study: emission directionality of a triangular photonic molecule}

In this section, we illustrate how the emission profile of CF states of PMs can differ from that of QB states. As a representative example, we consider a triangular PM composed of three circular photonic atoms of identical relative permittivity $\epsilon=4$ and radii $u$ arranged on the vertices of an equilateral triangle of side $2.5 u$. The signature of four QB states of this geometry, as well as the associated CF states, is shown in fig. \ref{fig:detmap}. The QB states are located in the frequency range $5.36 \leq k'u \leq 5.42$. For a value of the real exterior frequency $k=5.5$, the eigenfrequencies of the CF states (in the complex $K$ plane) are shifted towards lower frequencies as well, as can be seen from fig. \ref{fig:detmap}b. This situation where the real exterior frequency $k$ is shifted from $k'$ can occur if the central frequency of the gain transition is shifted as well. 

\begin{figure}
\centering
\includegraphics[]{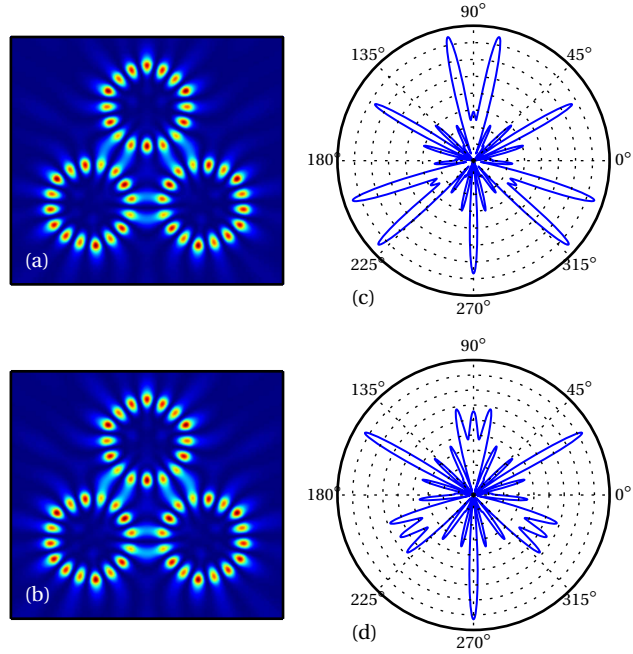}
\caption{(a) Amplitude profile of a QB state of a triangular PM. (b) Profile of the corresponding CF state. The position of this eigenstate is indicated by a white circle in fig. \ref{fig:detmap} (c-d) Comparison of the far-field profile $|g(\theta)|$ for each of the two kind of eigenstates. Arbitrary intensity units.}\label{fig:field}
\end{figure}

We have previously shown \cite{Gagnon2014} how shifts in the exterior frequency may affect the lasing thresholds of the PM and how the computation of CF states is necessary to precisely determine which of the modes will lase first. Another important characteristic of PM lasers which can be affected by such a shift is the emission profile. To illustrate this, we single out the eigenmode indicated by a white circle in fig. \ref{fig:detmap}. As can be seen in fig. \ref{fig:field}, the near-field profile of both the corresponding QB and CF states is rather similar, with the exception that the field of the CF state does not grow exponentially outside the PM. However, the strongest far-field emission direction of CF states does not correspond to that of the QB states. Indeed, the strongest directions are aligned with the triangle vertices in the case of the QB state, whereas they are aligned with the triangle edges in the case of the CF states. Moreover, the CF state far-field profile exhibits three sharper peaks, whereas there are six peaks in the QB state profile. This example shows the importance of using SALT for adequate characterization of directional emission.

\section{Summary and outlook}

In this contribution, we have characterized the emission profile of a triangular PM using the SALT theory. Specifically, we have shown that the privileged emission directions of the constant-flux states of this geometry differs if the exterior frequency $k$ is shifted from the natural QB eigenfrequency $k'$. Since the exact value of the exterior frequency $k$ depends on the gain medium parameters, such as the gain center frequency and its linewidth, this example shows that tuning the geometry of the cavity relative to those parameters may represent an additional control parameter to select different privileged emission directions. A more detailed parametric study will allow to fully characterize this interesting physical effect.

The authors acknowledge financial support from the Natural Sciences and Engineering Research Council of Canada (NSERC). J.D. and J.L.D. are grateful for a research fellowship from the Canada Excellence Research Chair in Photonic Innovations of Y. Messaddeq.

\end{document}